\documentclass[preprint,showpacs,preprintnumbers]{revtex4}
\usepackage{latexsym}
\usepackage{graphics}
\usepackage{epsf}
\usepackage{epsfig}
\usepackage{graphicx}
\usepackage{dcolumn}
\usepackage{bm}
\usepackage{latexsym}
\usepackage{booktabs}
\usepackage{amssymb}
\usepackage[english]{babel}
\usepackage{hyperref}

\newcommand {\be}{\begin{equation}}
\newcommand {\ee}{\end{equation}}
\newcommand {\bea}{\begin{eqnarray}}
\newcommand {\eea}{\end{eqnarray}}

\begin{document}

\noindent


\title{Kuramoto Oscillators on Chains, Rings and Cayley-trees}

\author{Filippo Radicchi and Hildegard Meyer-Ortmanns}
\affiliation{School of Engineering and Science,
International University Bremen,\\
P.O.Box 750561, D-28725 Bremen, Germany}

\date{\today}

\begin{abstract}
  We study systems of Kuramoto oscillators, driven by one pacemaker, on
  $d$-dimensional regular topologies like linear chains, rings,
  hypercubic lattices and Cayley-trees. For the special cases of
  next-neighbor and infinite-range interactions , we derive the
  analytical expressions for the common frequency in the case of
  phase-locked motion and for the critical frequency of the
  pacemaker, placed at an arbitrary position on the lattice, 
so that above the critical frequency no phase-locked motion is possible. These expressions
  depend on the number of oscillators, the type of coupling, the coupling strength,
  and the range of interactions. In particular we show that the
  mere change in topology from an open chain with free boundary conditions to a ring
  induces synchronization for a certain range of pacemaker
  frequencies and couplings, keeping the other parameters fixed.
  We also study numerically the phase evolution above the critical
  eigenfrequency of the pacemaker for arbitrary interaction ranges
  and find some interesting remnants to phase-locked motion below
  the critical frequency.
\end{abstract}

\pacs{05.45.Xt, 05.70.Fh}

\maketitle
\section{Introduction}
\label{sec:introduction} Synchronization is an ubiquitous
phenomenon, found in a variety of natural systems like fireflies,
where the males flash light in synchrony to attract the females
\cite{fireflies}, chirping crickets \cite{crickets}, or like
neural systems, but also utilized for artificial systems of
information science in order to enable a well-coordinated behavior
in time. As an important special case, the coordinated behavior
refers to similar or even identical units like oscillators that
are individually characterized by their phases and amplitudes. A
further reduction in the description was proposed by Kuramoto for
an ensemble of $N$ oscillators \cite{kuraoriginal}, after Winfree
had started with a first model of coupled  oscillators
\cite{winfree}. Within a perturbative approach Kuramoto showed
that for any system of weakly coupled and nearly identical
limit-cycle oscillators, the long-term dynamics is described by
differential equations just for the phases $\varphi_i$ (not for
the amplitudes) with mutual interactions, depending on the phase
differences in a bounded form. Kuramoto solved this model for the
special case of all-to-all coupled oscillators, for identical
interaction functions, and for eigenfrequencies, taken out of a
narrow distribution. It is, however, this model, later named after
him, that nowadays plays the role of a paradigm for weakly and
continuously interacting oscillators.  In a large number of
succeeding publications the original Kuramoto model was
generalized in various directions, for a recent review see
\cite{kuramotoreview}. In particular the eigenfrequencies were
specialized in a way that one oscillator plays the role of a
pacemaker with eigenfrequency different from zero, while all
others have eigenfrequency zero \cite{yamada}. In \cite{yamada}
the pacemaker was placed at a specific position, the beginning of
a linear chain. Moreover the interaction range was changed from
all-to-all to next neighbor interactions
\cite{daido}\cite{strogatz}. In \cite{daido} the system was
studied numerically, and an abrupt change from long-to short-term
interaction was suggested.
\\
In this paper we consider a system of  Kuramoto oscillators, coupled on
various regular lattice topologies, and driven by a pacemaker,
placed at an arbitrary site of the lattice. The interaction range
is neither restricted to next-neighbor nor to all-to-all
couplings, but varied by a parameter $\alpha$ that tunes the decay
with the distance between the interacting units. We analytically
derive the common frequency $\Omega$ of phase-locked motion and
the upper bound on the absolute value of the ratio of the pacemaker's frequency to the
coupling strength $\left|\omega_s/K\right|$ for next-neighbor and all-to-all
couplings in case of $d$-dimensional regular lattices with open or
closed boundaries, in particular for a ring and an
open chain. The results show the dependence on the size of the
system (total number $N$ of oscillators), the position $s$ of the
pacemaker (for open boundary conditions) and the (a)symmetric
treatment of the pacemaker's interaction with the other Kuramoto
oscillators (tuned by the parameter $\epsilon_s$). The intermediate
interaction range $(0<\alpha<\infty)$ as well as the phase
evolution in the desynchronized phase (above the threshold for
phase-locked motion) are treated numerically.

\vspace{0.2cm}

The paper is organized as follows. In section II we define the
model and identify the various special cases. Section
\ref{sec:phase-locked} is devoted to the study of phase-locked
motion. We calculate the common frequency and the critical
threshold for phase-locked motion for a linear chain and
$\alpha\to\infty$ (section \ref{sec:linear_chain}), for a ring and
$\alpha\to\infty$ in section \ref{sec:ring}, the case of
all-to-all couplings ($\alpha=0$) in section \ref{sec:all-to-all}.
Our main result about a switch to synchronization via the closure
of a chain is stated in section \ref{sec:switch_sync}. Section
\ref{sec:intermediate} presents our results of numerical
simulations for the intermediate interaction range
($0<\alpha<\infty$). In section \ref{sec:higher_dim} we generalize
the previous results to regular lattice topologies in higher
dimensions, i.e. to Cayley-trees and d-dimensional hypercubic
lattices). In section \ref{sec:above_critical} we report on
numerical results for the phase portrait above the critical
threshold, in the desynchronized phase, in which the phases appear
as a superposition of a fast motion, determined by the pacemaker,
and a slow motion, determined by the interaction of all other
oscillators. Finally we give the summary and conclusions in
section \ref{sec:conclusions}. Details of the analytical
calculations are postponed to the Appendix.

\section{The Model}
\label{sec:model}

\begin{figure}[ht]
\includegraphics*[width=0.97\textwidth]{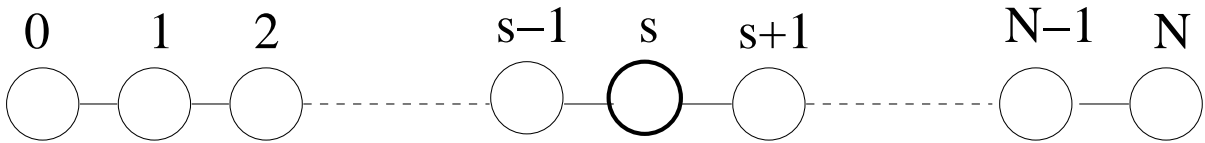}
\caption{One-dimensional lattice with $\left(N+1\right)$ Kuramoto oscillators, labelled by their coordinates.} \label{fig:linear_chain}
\end{figure}

The system is defined on a $d$-dimensional lattice, where each
site is a limit-cycle oscillator. The phase of the $i$-th
oscillator obeys the dynamics determined by
\begin{equation}
\dot{\varphi}_i = \delta_{i,s} \omega_s +
(1+\delta_{i,s}\epsilon_s) \ \frac{K}{\eta_i} \ \sum_{j\neq i}
 \frac{\sin{(\varphi_j-\varphi_i)}}{r_{ji}^\alpha}  \;\;\; , \; \forall \; i=0,\ldots ,N \;\;\; ,
\label{def:model}
\end{equation}
where
\begin{equation}
\eta_i := \sum_{j\neq i} r^{-\alpha}_{ji}\;\;\; , \; \forall \; i=0,\ldots ,N \;\;\; . \label{def:eta}
\end{equation}
In particular  $\varphi_i$ is the phase of the i-th oscillator,
the index $s$ labels the pacemaker with eigenfrequency $\omega_s
\neq 0$, while all other oscillators have eigenfrequency zero. The
parameter $-1\leq \epsilon_s \leq 0$ serves to tune the
interaction of the pacemaker with the other oscillators. For
$\epsilon_s=0$ the pacemaker is on the same footing as the other
oscillators, being only distinct by its eigenfrequency $\omega_s$.
For $\epsilon_s=-1$ its interaction is asymmetric in the sense
that the pacemaker  influences the other oscillators, but not vice
versa (the pacemaker acts as an external force), so that it is on
a different level in the hierarchy of the system. (In natural
systems both extreme cases as well as intermediate couplings may
be realized: a pacemaker being one out of an ensemble of
oscillators or being distinct by additional features to the
eigenfrequency.) The constant $K>0$ parameterizes the coupling
strength, $\eta_i$ represents a space-dependent normalization, it
replaces the factor $N$ that is used for all-to-all interactions
and turns out to be essential for the $\alpha$-dependence of the
synchronization threshold, cf. section \ref{sec:intermediate}. The
phases of the i-th and j-th oscillators interact via the sine
function (as originally proposed by Kuramoto), of which two
properties enter the analytical predictions: the sine is bounded
and an odd function. The distance $r_{ij}$, given in units of the
edges of the lattice, enters to the power $-\alpha$, $\alpha=0$
gives the space-independent all-to-all coupling, originally
considered by Kuramoto, $\alpha\to\infty$ is understood to
restrict the interactions to nearest neighbors. The label of the
oscillator simultaneously specifies the location in space, see
e.g. Figure \ref{fig:linear_chain} for a one-dimensional chain,
where coordinate $s$ specifies the position of the pacemaker.

System (\ref{def:model}) contains the original model of Kuramoto as
special case for $\epsilon_s=0$, $\alpha=0$ and absent
$\delta_{i,s}$ (all oscillators with an eigenfrequency chosen at
random from a probability distribution). Note that the
interaction strength is symmetric between oscillators $i,j$, with
$i,j\neq s$, continuous in time and instantaneous. It depends on
the natural or artificial systems that are modelled by Kuramoto
oscillators, whether these features amount to appropriate
idealizations or to an oversimplification. For example, the
interactions between fireflies propagate with the very velocity of
light. In view of their mutual distances on the trees the delay
caused by the finite velocity of light should be completely
negligible. When the interaction proceeds via pulses as in neural
systems, the continuous interaction [as in system (\ref{def:model})]
is also not adequate.

\section{Phase-locked Motion}
\label{sec:phase-locked} In this section we consider the
conditions for having phase-locked motion, in which the phase
differences between any pair of oscillators remain constant over
time, after an initial relaxation time. In general two quantities
are of interest for phase-locked motion, the frequency $\Omega$,
in common to all oscillators in the phase-locked state, and the
critical threshold in terms of the absolute value of the eigenfrequency of the pacemaker $\left| \omega_s \right|$
over the coupling strength $K$, i.e. $\left|\omega_s/K\right|_C$, that gives
an upper bound for having phase-locked motion as a stable state.
We analytically calculate (see the Appendix) $\Omega$
for any dimension $d$ of the lattice and for arbitrary range of the
interaction (i.e. $0  \leq \alpha \leq \infty$). Imposing the
phase-locked condition
\[
\dot{\varphi}_i \equiv \Omega \;\;\; , \; \forall \; i=0,\ldots ,N
\]
to system (\ref{def:model}) and using the fact that the sine is an odd
function, one easily obtains
\begin{equation}
\Omega\;=\; \frac{N\eta_s\omega_s}{(1+N\epsilon_s)\sum_{i \neq s}
\eta_i + \sum_{j \neq s} \sum_{i \neq j} \eta_i} \; \; ,
\label{eq:common_frequency}
\end{equation}
where the $\eta_i$s are defined in Eq.(\ref{def:eta}). They
contain the informations about the range of the interactions and
about the topology on which the system is defined (i.e. $\eta_i$
is a function of $\alpha$, it depends on the dimension $d$ and on
the boundary conditions of the lattice). Moreover,  in case of
periodic boundary conditions and in case of long-range
interactions ($\alpha=0$), we find
\[
\eta_i \equiv \eta \;\;\; , \; \forall \; i=0,\ldots, N \;\;\; ,
\]
or, in other words, the normalization factor defined in
Eq.(\ref{def:eta}) becomes independent on $i$ (i.e. the position
on the lattice of the $i$-th oscillator). In this case we are able
to prove that Eq.(\ref{eq:common_frequency}) takes the explicit
form
\begin{equation}
\Omega = \frac{\omega}{N+\epsilon N+1}
\;\;\;.\label{eq:common_frequency_ring}\end{equation}

The main result is that Eq.(\ref{eq:common_frequency_ring}) is
independent on $\alpha$. Moreover, in such cases the position of
the pacemaker  does not change the behavior of the system.
Therefore  we dropped the index $s$ of the pacemaker in
Eq.(\ref{eq:common_frequency_ring}) .
\\
As a next step, we calculate the critical threshold
$\left|\omega_s/K\right|_C$ for a one-dimensional lattice and
different range of interactions: analytically for next-neighbor
$(\alpha\to\infty)$ and for long-range $(\alpha=0)$ interactions
(section \ref{sec:linear_chain}-\ref{sec:all-to-all}), numerically
for intermediate coupling range (section \ref{sec:intermediate}).
Both quantities, the common frequency $\Omega$ and the critical
threshold $\left|\omega_s/K\right|_C$, have an analytical
generalization to higher dimensions $d>1$ (section
\ref{sec:higher_dim}).
\\
All numerical results (apart from those of Figure \ref{fig:average_frequency}) are obtained with an accuracy of $5 \cdot 10^{-6}$ by
integrating the system (\ref{def:model}) by the fourth-order
Runge-Kutta method (with fixed time step $\Delta t = 0.05$).

\subsection{Kuramoto oscillators on a linear chain with next-neighbor
interactions} \label{sec:linear_chain}

We consider $(N+1)$ Kuramoto oscillators coupled along a chain as
indicated in Figure \ref{fig:linear_chain} with free boundary
conditions at the positions $0$ and $N$, and the pacemaker located
at position $s$, $0\leq s\leq N$. For this simple topology,
Eq.(\ref{def:eta}) takes the form
\begin{equation}
\eta_i = \theta(i) \sum_{j=1}^i j^{-\alpha} \;+\; \theta(N-i)
\sum_{j=1}^{N-i} j^{-\alpha} \; , \forall i=0,\ldots,N \;
,\label{eq:eta_chain}
\end{equation}
where $\theta(x)$ is the Heaviside step function [$\theta(x)=1$
for $x>0$ and $\theta(x)=0$ for $x\leq 0$].
Eq.(\ref{eq:eta_chain}) inserted into
Eq.(\ref{eq:common_frequency}) gives the proper value of the
common frequency $\Omega$ as function of the position $s$ and the
eigenfrequency $\omega_s$ of the pacemaker, the size of the system
$N$ as well as the range of the interaction, parameterized by
$\alpha$.
\\
To derive the critical frequency of the pacemaker in the case of
next-neighbor interactions $(\alpha\to\infty)$, we  use the odd
parity of the sine function. We derive a recursive relation
between the phase differences of next-neighbor oscillators. Using
the boundary conditions it is possible to express these recursive
equations in terms of $s$, $N$, $\Omega$, and $K$. On the other
hand, the sine function is bounded. These bounds induce the bound
on the maximal pacemaker frequency $\left|\omega_s/K\right|_C$ that is
compatible with a phase-locked motion. During the recursions
between phase differences we have to distinguish whether we move
to the boundary to the right or to the left of the pacemaker $s$,
placed between $0 < s < N$; either the boundary conditions
enter for oscillator $N$ or oscillator $0$. Thus if we move to the
right, we obtain for the maximal pacemaker's
frequency
\begin{equation}
^R \left| \frac{\omega_s}{K}\right|_C =
\frac{(N-1)(1+\epsilon_s)+1}{2N-2s-1} \;, \label{eq:critic_right}
\end{equation}
so that all oscillators to the right ($R$) of  $s$ approach a
phase-locked state, and
\begin{equation}
^L \left|\frac{\omega_s}{K}\right|_C =
\frac{(N-1)(1+\epsilon_s)+1}{2s-1} \label{eq:critic_left}
\end{equation}
as bound for oscillators on the left ($L$) of $s$ to synchronize in a
phase-locked motion. Since we are interested in a state with all
oscillators of the chain being phase-entrained, we need the
stronger condition given by
\begin{equation}
\left|\frac{\omega_s}{K}\right|_C = \min{\left[ \; ^R
\left|\frac{\omega_s}{K}\right|_C \; , \;  ^L\left|
\frac{\omega_s}{K}\right|_C \; \right]}  \; \; . \label{eq:critic}
\end{equation}
For the pacemaker placed at the boundaries $s=N$ or $s=0$ we have
\begin{equation}
\left| \frac{\omega_s}{K} \right|_C = \frac{2N +2N\epsilon_s
-\epsilon_s}{2N-1} \; \; . \label{eq:critic_bound}
\end{equation}
If the pacemaker is located in the middle of the chain, i.e. at
$s=N/2$ for $N$ even or $s=(N\pm 1)/2$ for $N$ odd, we have
\begin{equation}
\left| \frac{\omega_{N/2}}{K} \right|_C =
\frac{(N-1)(1+\epsilon_{N/2})+1}{N-1} \textrm{ , if } N \textrm{
is even} \label{eq:critic_middle_even}
\end{equation}
and
\begin{equation}
\left| \frac{\omega_{(N\pm 1)/2}}{K} \right|_C =
\frac{(N-1)(1+\epsilon_{(N\pm 1)/2})+1}{N} \textrm{ , if } N
\textrm{ is odd} \;. \label{eq:critic_middle_odd}
\end{equation}
Note that for $\epsilon_s=-1$ the
critical ratio goes to zero for $N\to\infty$, while it approaches
one for $\epsilon_s=0$ and $N\to\infty$, but the common frequency $\Omega$
goes to zero in this case for $N\to\infty$. Moreover we see the
stronger the coupling $K$, the larger $\left|\omega_s\right|$ may be up to
which the system is able to follow the pacemaker.

\begin{figure}[ht]
\includegraphics*[width=0.97\textwidth]{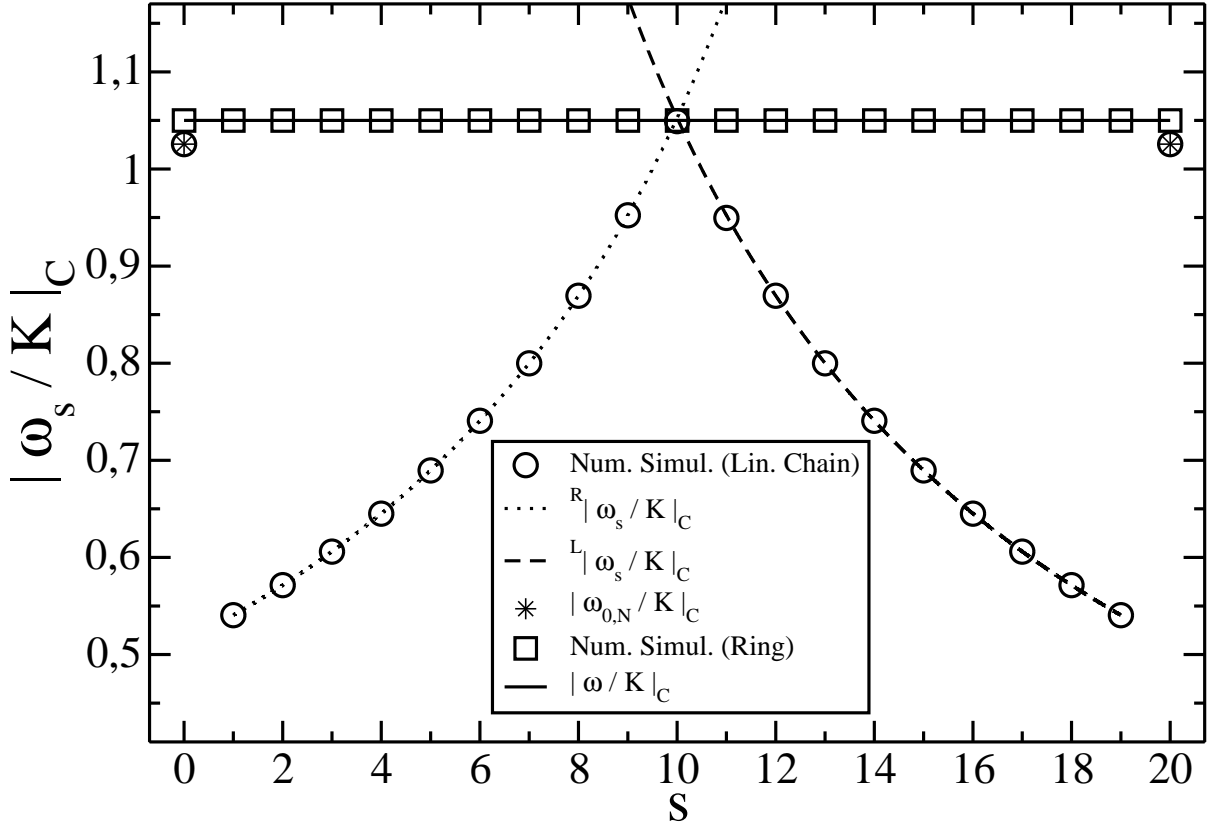}
\caption{Critical threshold $\left|\omega_s/K\right|_C$ as
function of the position $s$ of the pacemaker on a one-dimensional
lattice (for both free and periodic boundary conditions). Here
$N=20$, $\alpha=100.0$ and $\epsilon_s=0$. Theoretical predictions
[dotted, dashed and full lines from Eq.s(\ref{eq:critic_right}),
(\ref{eq:critic_left}) and (\ref{eq:critic_ring}),respectively,
and stars from Eq.(\ref{eq:critic_bound})] are in perfect
agreement with results obtain from numerical simulations (circles
and squares).} \label{fig:critic_one-dim}
\end{figure}

\subsection{Ring topology and next-neighbor interactions}
\label{sec:ring} For $\alpha\to\infty$ and $(N+1)$ oscillators
placed on a ring, it is easy to see  that the normalization factor
defined in Eq.(\ref{def:eta}) is the same for all oscillators $i$
[therefore Eq.(\ref{eq:common_frequency_ring}) gives the right
value of $\Omega$]. It can be written as

\begin{equation}
\eta= \left\{ \begin{array}{ll} 2 \sum_{j=1}^{N/2} \; j^{-\alpha}
& \textrm{ , if } N \textrm{ is even}
\\
2 \sum_{j=1}^{(N-1)/2} \; j^{-\alpha} \; + \; \left( \frac{N+1}{2}
\right)^{-\alpha}  & \textrm{ , if } N \textrm{ is odd}
\end{array}
\right. \;\;\; . \label{eq:eta_ring}
\end{equation}

The derivation on the upper bound of the pacemaker's frequency
$\left|\omega/K\right|_C$ for a ring proceeds in analogy to Eq.s
(\ref{eq:critic_right}), (\ref{eq:critic_left}) and
(\ref{eq:critic_bound}), apart from the fact that the boundary
conditions are periodic
\begin{equation}
\varphi_t \equiv \varphi_{\left(N+1\right)+t} \;\;\; , \; t \in
\mathbb{Z}\;. \label{eq:periodic_conditions}
\end{equation}
The result is
\begin{equation}
\left| \frac{\omega}{K}\right|_C\;=\;\frac{N+N\epsilon+1}{N}\;\;\;.
\label{eq:critic_ring}
\end{equation}

\subsection{All-to-all coupling}
\label{sec:all-to-all}
In case of all-to-all coupling
($\alpha=0$) the topology of the system has no longer an effect on
the phase entrainment; moreover the normalization factor of
definition (\ref{def:eta}) is $\eta_i \equiv N \;\; , \; \forall \; i=0,\ldots , N$, therefore the common
frequency is given by Eq.(\ref{eq:common_frequency_ring}). Since
we know that in the case of all-to-all couplings the mean-field
approximation becomes exact, it is natural to introduce a quantity
that usually serves as order parameter \cite{kuraoriginal} for indicating the
synchronized phase, i.e.
\begin{equation}
R\;e^{i\psi}:= \frac{1}{N} \sum_{j \neq s} e^{i\varphi_j} \;\;\; .
\label{def:order_parameter}
\end{equation}
In our definition $R\;e^{i\psi}$ differs from the usual order
parameter only by the fact that the sum runs over $j\neq s$.
In terms of the order parameter we can rewrite the system
(\ref{def:model}) according to
\begin{equation}
\left\{
\begin{array}{ll}
\Omega=\omega_s+(1+\epsilon_s)KR \sin{(\psi-\varphi_s)} &
\\
\Omega=
\frac{K}{N}\sin{(\varphi_s-\varphi_i)}+KR\sin{(\psi-\varphi_i)} &
\textrm{ , } \forall \; i\neq s
\end{array}
\right. \; . \label{eq:all-to-all1}
\end{equation}
As we see from Eq.s (\ref{eq:all-to-all1}), all oscillators
different from the pacemaker satisfy exactly the same equation.
One possible solution is given by

\[
\varphi_i \equiv \psi  \; \textrm{ , } \forall \; i \neq s \; \; \; ,
\]

so that all phases of the oscillators different from the pacemaker
are equal. Using system (\ref{eq:all-to-all1}) and the fact that
the sine function is bounded, it is easily seen that the critical
ratio $\left|\omega/K\right|_C$ is still given by
Eq.(\ref{eq:critic_ring}), valid for the ring and
$\alpha\to\infty$.

\subsection{A switch to synchronization}
\label{sec:switch_sync} 
\begin{figure}[ht]
\includegraphics*[width=0.97\textwidth]{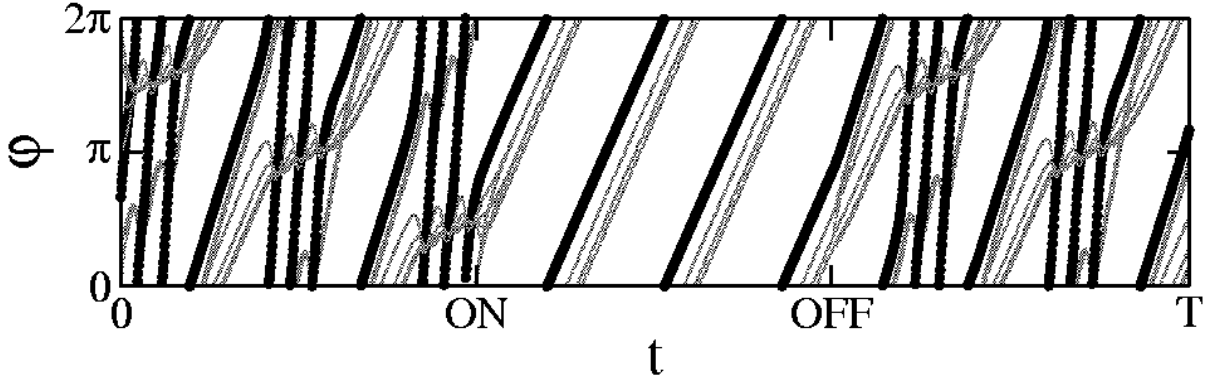}
\caption{Phase portrait of a system of Kuramoto oscillators on a
one-dimensional lattice with $N=6$, $\alpha=5$, $s=2$,
$\epsilon_s=0$ and $\omega_s/K=1$. The phase of the pacemaker is
represented by the fat black line, while the phases of the other
oscillators are drawn with thin dark-grey lines. We start from a
chain (free boundary conditions), for which we are above the
critical threshold given in Eq.(\ref{eq:critic}), here the stable
state of the system is characterized by a non-constant frequency
(the frequency of the oscillators are different from each other
and vary during time). We observed the system for $8000$
integration steps altogether. At the instant ``ON'' ($T/3$) we
closed the chain, passing to a ring topology: now the ratio
$\omega/K$ is less than the one given by
Eq.(\ref{eq:critic_ring}), so that the system can synchronize in
the frequency domain (it reaches a phase-locked motion where all
the phases have the same constant slope). At instant ``OFF''
($2T/3$) we opened the ring and the system falls again into a
non-phase-locked motion, similarly to that seen in the interval
[$0$,ON]. }\label{fig:switch}
\end{figure}

Let us summarize the results we obtained
so far in Figure \ref{fig:critic_one-dim}. The full lines
represent the analytical results for $\left|\omega_s/K\right|_C$
for the open chain with $\alpha\to\infty$, more precisely for
$\alpha=100$, and the ring for  $\alpha=100$, the circles and
squares represent numerical data that reproduce the analytical
predictions within the numerical accuracy. All results are
obtained for $N=20$ and $\epsilon_s=0$, they  are plotted as a
function of the pacemaker's position that only matters in case of
the open chain and for $\alpha > 0$. The horizontal line at
$\left|\omega_s/K\right|_C=(N+1)/N$ obviously refers to the ring,
the two branches (left and right), obtained for the chain, cross
this line when the pacemaker is placed at $s=N/2$ in the middle of
the chain. When the pacemaker is located at the boundaries $s=0$
and $s=N$, we obtain two isolated data points, cf. the
corresponding
Eq.(\ref{eq:critic_bound}), close to the horizontal line.\\
Let us imagine that for given $N$ and $\epsilon_s$ the absolute value of the pacemaker's
frequency $\left|\omega_s\right|$ and the coupling $K$ are specified out of a
range for which
\[
\left|\frac{\omega_s}{K}\right|_C \; < \; \left| \frac{\omega_s}{K}\right| \;
\leq \; \left| \frac{\omega}{K} \right|_C
\]
such that the ratio is too large to allow for phase-locked motion
on a chain, but small enough to allow the phase-entrainment on a
ring. It is then the mere closure of the open chain to a ring that
leads from non-synchronized to synchronized oscillators  with
phase-locked motion. Therefore, for a whole range of ratios
$\left|\omega_s/K\right|$, no fine-tuning is needed to switch to a
synchronized state, but just a simple change in topology. Because
of its simplicity we suspect that this mechanism is realized in
biological systems that need the possibility of an easy switch
between synchronization and desynchronization. In our numerical
integration we simulated  such a switch and plot the phase
portrait in Figure \ref{fig:switch}. The phase evolution $\varphi$
as function of time is always projected to the interval
$\left[0,2\pi \right)$: we use a thick black line for the phase of
the pacemaker and thin dark-grey lines for the other oscillators.
In the numerical simulation with $T=8000$ integration steps over
time we analyzed a one-dimensional lattice of Kuramoto oscillators
with  $N=6$, $\alpha=5.0$, $s=2$, $\omega_s/K=1$ and
$\epsilon_s=0$. In the time interval from $0$ to ``ON'' ($T/3$) we
see for the system motions with non-constant slopes: more steep
for the pacemaker and the left part of the system (oscillators
$i=0,1$) and  less steep for the right part (oscillators
$i=3,4,5,6$). At the instant ``ON'' we close the chain, passing to
a ring topology. The system almost instantaneously reaches a
phase-locked motion (all phases moving with the same slope). At
time ``OFF'' ($2T/3$) we open the ring, again, and the system
shows qualitatively the same behavior as before the closure.

Furthermore it should be noticed from Figure
\ref{fig:critic_one-dim} that it is also favorable to put the
pacemaker at the boundaries of an open chain to facilitate
synchronization. For $d=1$ the pacemaker then has to entrain only
one rather than two nearest neighbors so that the range of allowed
frequencies $\left|\omega_s\right|$ increases.
\\
>From these results it is natural to try to utilize the
topology in a way that facilitates synchronization in artificial
networks when synchronized states are needed. Given $N$
oscillators and a number of $p_0$ pacemakers, one can optimize the
placement of the pacemakers when only a limited range of couplings
is available and no other fine-tuning of parameters is feasible.

\subsection{Intermediate range of couplings}
\label{sec:intermediate} 
\begin{figure}[ht]
\includegraphics*[width=0.97\textwidth]{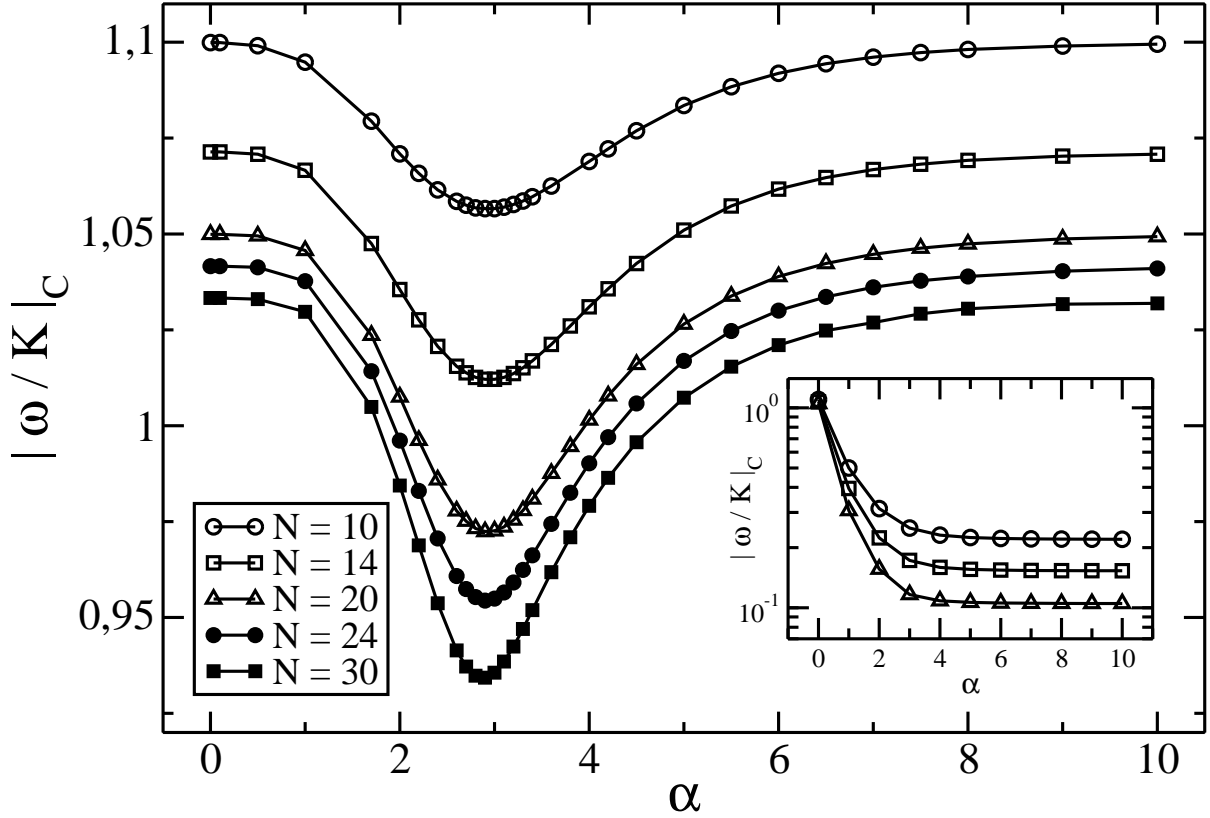}
\caption{Critical ratio $\left| \omega / K \right|_C$ for a
one-dimensional ring with $\epsilon=0$ as function of $\alpha$ for
various system sizes, showing a non-monotonic behavior with a
minimum at $\alpha_M \leq 3$.  The inset shows the same critical
ratio for $N=10,14,20$, but for the system (\ref{def:model}) with
$\eta=N$, chosen independently on $\alpha$: now
$\left|\omega/K\right|_C$ behaves as a monotonically
 decreasing function of $\alpha$.} \label{fig:alpha_ring}
\end{figure}

\begin{figure}[ht]
\includegraphics*[width=0.97\textwidth]{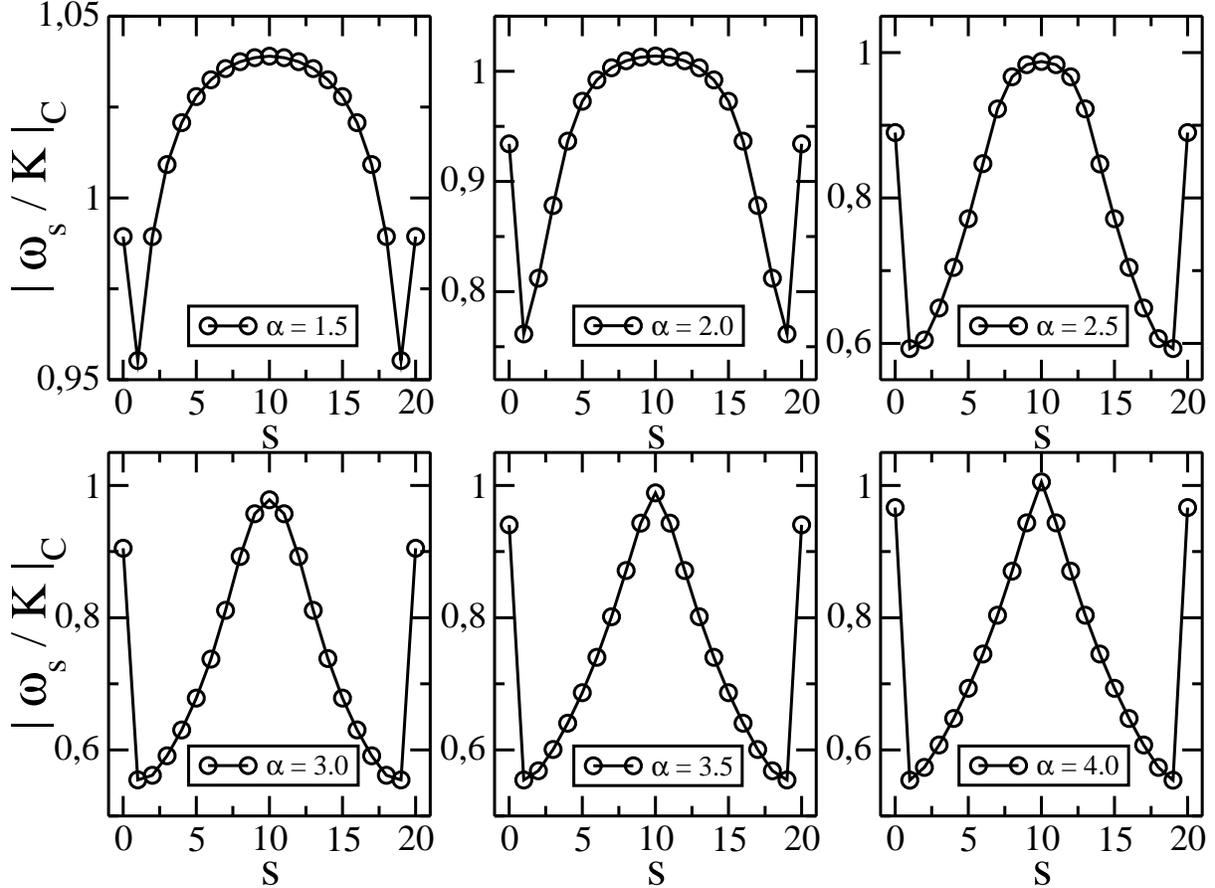}
\caption{Critical ratio $\left| \omega_s /K \right|_C$ as function
of the pacemaker's position for an open chain and various values
of $\alpha$, indicating a qualitative change in the behaviour
between $\alpha=2.5$ and $\alpha=3$.} \label{fig:alpha_chain}
\end{figure}

Recall from section
\ref{sec:phase-locked} that the common frequency $\Omega$ of
phase-locked motion was independent on $\alpha$ in case of a ring
and depending on $\alpha$ in a well-defined way [cf.
Eq.s(\ref{eq:common_frequency}),(\ref{eq:eta_chain})] in case of
an open chain. The upper bound on the pacemaker's frequency,
however, was analytically derived only for $\alpha=0$ and
$\alpha\to\infty$. For intermediate values of $\alpha$ we studied
the behavior of $\left|\omega_s/K\right|_C$ as a function of
$\alpha$ numerically. The results are shown in Figures
\ref{fig:alpha_ring} and \ref{fig:alpha_chain}.

In case of the ring topology (Figure \ref{fig:alpha_ring}) we
observe a non-monotonic function of $\alpha$ with a minimum at
$\alpha_M \leq 3$ that seems to decrease as the system size
increases, here from $N=10$ to $30$. The shape of
$\left|\omega_s/K\right|_C$ depends on the choice of the
normalization with $K/\eta_i$ in the interaction term of
Eq.(\ref{def:model}). If we had chosen $K/N$ instead of
$K/\eta_i$, we would find a monotonic decrease (inset of Figure
\ref{fig:alpha_ring}). Instead we see here that all-to-all and
next-neighbor interactions lead to the same ratio of
$\left|\omega/K\right|_C$, while synchronization becomes more and
more difficult for $\alpha$ around $\alpha_M$. The existence of
$\alpha_M$ is also supported  by numerical results of
\cite{rogers} for $N=100$ oscillators, where it was found that the
system approaches the same behavior as the original Kuramoto model
with $\alpha=0$ for $\alpha\leq 2$ so that there is a different
behavior for long-  and intermediate-range couplings. We expect
that our values for $\alpha_M$ decrease to $\alpha_M\propto 2$ for
larger system sizes.
\\
Figure \ref{fig:alpha_chain} shows the same critical ratio
$\left|\omega_s/K\right|_C$ as a function of the pacemaker's
position $s$ for an open chain, $N=20$, $\epsilon_s=0$, and
various values of $\alpha$. Here it is interesting to see that the
shape of $\left|\omega_S/K\right|_C$ drastically changes around
the same value of $\alpha_M\leq 3$ as for the ring: for
$\alpha<\alpha_M$, the critical fraction looks differentiable for
all positions $s$ of the pacemaker along the chain, apart from the
boundaries $s=0$ and $s=N$, while for $\alpha>\alpha_M$ it appears
non-differentiable around $s=N/2$ and has a cusp. An analytical
understanding of $\left|\omega_S/K\right|_C$ as a function of $s$
and $\alpha$ is still missing.

\subsection{Kuramoto oscillators on regular structures in higher dimensions}
\label{sec:higher_dim}

\begin{figure}[ht]
\includegraphics*[width=0.97\textwidth]{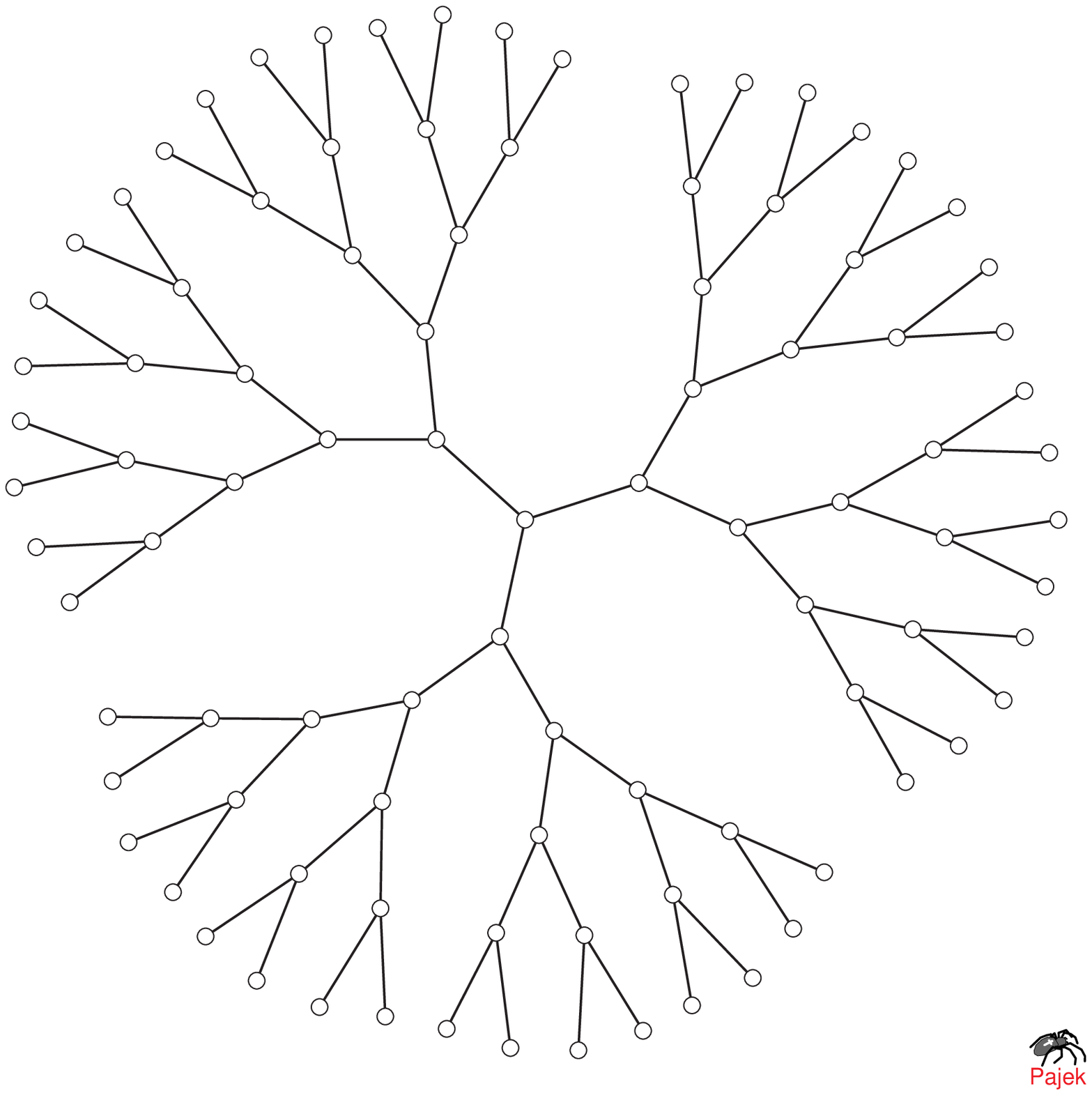}
\caption{A Cayley tree with $z=3$ branches and radius $R=5$.}
\label{fig:cayley}
\end{figure}

\begin{figure}
\includegraphics[width=0.97\textwidth]{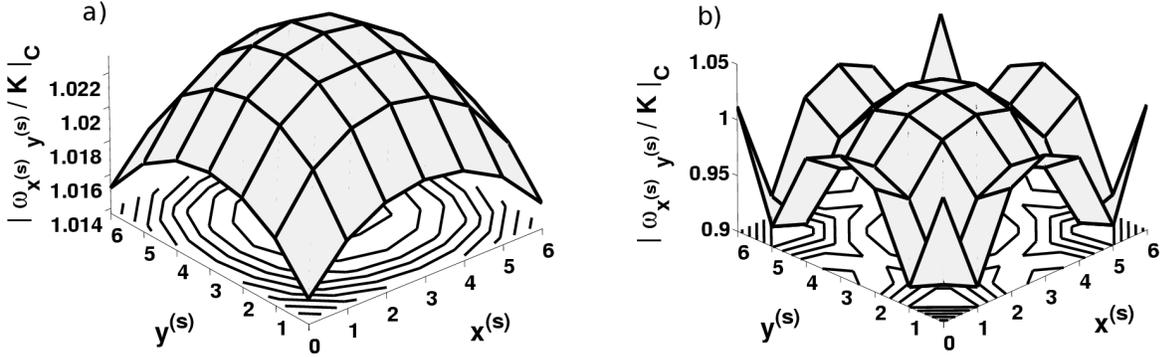}
\caption{Critical ratio $\left|\omega_{x^{(s)}
y^{(s)}}/K\right|_C$ for a two-dimensional lattice with $N=6$,
$\epsilon_{s}$=0 and a) $\alpha=1.0$, b) $\alpha=100.0$.}
\label{fig:2dim}
\end{figure}

In order to generalize our previous results to regular structures
in higher dimensions, we consider Cayley-trees and hypercubic
lattices. Cayley-trees have $z$ branches to nearest neighbors at
each node, cf. Figure \ref{fig:cayley}, apart from nodes in the
outermost shell, where the number of nearest neighbors is only
$z-1$. Cayley-trees generalize linear chains from $z=2$ to $z>2$
branchings at each node. They provide a geometry that often allows
for exact analytical solutions, cf. the percolation problem on a
Cayley-tree \cite{havlin}. Cayley-trees are infinite-dimensional
in the sense that the number of nodes in the boundary (outermost
shell at distance $R$ from the origin) are proportional to the
volume, i.e. the total number of nodes in the tree. Throughout our
calculations we place one pacemaker in the center of the tree and
assign all other Kuramoto oscillators with eigenfrequency zero to
the remaining nodes. We start from a modified version of the
equations (\ref{def:model})

\begin{equation}
\dot{\varphi}_i = \delta_{i,0} \omega_0 +
 \frac{K}{z} \ \sum_{j\neq i}
 A_{ji} \; \sin{(\varphi_j-\varphi_i)}  \;\;\; , \; \forall \; i=0,\ldots ,N \;\;\; ,
\label{def:cayley}
\end{equation}

where $A_{ji}$ is the adjacency matrix of a Cayley-tree. Naively
we would expect that the larger the number of branches  $z$ and
the larger the radius $R$, the more difficult gets the
synchronization of the tree. Therefore we calculate the common
frequency $\Omega$ and the critical threshold
$\left|\omega_0/K\right|_C$ as a function of $z$ and $R$. It is
convenient to assign  the  distance $r$ from the origin (center
node) as coordinate of a node  together with the path, along which
it is reached from the origin, that is $(r, s_1,s_2,...,s_r)$ with
$s_1=1,..,z$, $s_i=1,...,z-1$ for all shells $i=2,...,r$. Imposing
the condition of phase-locked motion, we see that all oscillators
at distance $r$ satisfy the same equation, independently on the
position within this shell (i.e. independently on the path along
which they are reached from the center), so that we finally can
drop $s_1,...,s_r$ and keep only the distance,
$\varphi_{r,s_1,s_2,...,s_r}\equiv\varphi_r$. Again one can derive
recursive relations for phase differences between phases $\varphi$
in neighboring shells (cf. the Appendix) and express the final
difference between shell $0$ and shell $1$ exclusively in terms of
the parameters of the system, i.e. $z$ and $R$, leading to
\begin{equation}
\Omega \; = \; \omega_0 \; \left( \ 1 \ + \ z \ \sum_{q=0}^{R-2} \
(z-1)^q \; + \; (z-1)^{R-1} \; \right)^{-1} \; \; \; .
\label{eq:common_frequency_cayley}
\end{equation}
As it is seen from Eq.(\ref{eq:common_frequency_cayley}), $\Omega$ goes to zero for $R\to\infty$, $z\to\infty$ or both. The critical threshold $\left|\omega_0/K\right|_C$
is obtained from an expression for $\sin(\varphi_r-\varphi_{r-1})$
(cf. the Appendix) that is always negative and monotonically
increasing as a function of $r$ if $\omega_0>0$ (or  always positive and monotonically
decreasing as a function of $r$ if $\omega_0<0$) , so that it takes a minimum (maximum) at
$r=1$. It is then again the bound of the sine function that leads to the critical
ratio
\begin{equation}
\left|\frac{\omega_0}{K}\right|_C \; = \; \frac{1 \ + \ z \
\sum_{s=0}^{R-2} \ (z-1)^s \ + \ (z-1)^{R-1} } {z \
\sum_{s=0}^{R-2} \ (z-1)^s \ + \ (z-1)^{R-1}} \; \; \; \; .
\label{eq:critic_cayley}
\end{equation}
Similarly to the result for $\left|\omega_s/K\right|_C$ for the linear chain,
this ratio goes to one for $z\to\infty$ or $R\to\infty$ or both,
while $\Omega\to 0$ in the same limiting cases. For the special
case of $z=2$ the Cayley-tree reduces to a linear open chain with
the pacemaker placed in the middle of the chain, and Eq.
(\ref{eq:critic_cayley}) reduces to Eq.(\ref{eq:critic_middle_even}) with  $\epsilon_s=0$.
\\
Since we are interested in finite systems, we also consider
Eq.(\ref{eq:critic_cayley}) as a condition on the critical size in
terms of $z$ and  $R$ which may not be exceeded for keeping
phase-locked motion if the pacemaker's frequency $\omega_0$ and
the coupling strength $K$ are given. The fact that for
sufficiently small $z$ and $R$ synchronization on a Cayley- tree
is possible, is in qualitative agreement with a result, recently
obtained for an ensemble of R\"ossler oscillators on a Cayley-tree
\cite{soonhmo}. R\"ossler oscillators that are individually even
chaotic systems, also approach a synchronized state on a
Cayley-tree if $z$ and $R$ are sufficiently small. \vspace{0.2cm}
We can extend qualitatively all the results so far obtained for
any dimension $d$, when system (\ref{def:model}) is placed on  a
hypercubic lattice with $(N_j+1)$ oscillators in each dimension,
so that we have an ensemble of $\Pi_{j=1}^d (N_j+1)$ Kuramoto
oscillators. The $i$-th oscillator's position is labelled by
$d$-dimensional vectors $\vec{x}^{(i)}$, with $0\leq x^{(i)}_j\leq
N\;\;\; , \; \forall \;j=1,\ldots ,d$. A phase-locked motion now
occurs for $\left|\omega_{x^{(s)}_1\ldots x^{(s)}_d}/K\right|$
$\leq \left| (\omega_{x^{(s)}_1 \ldots x^{(s)}_d}/K\right|_C$, the
critical ratio for the pacemaker's frequency at position
$\vec{x}^{(s)}=(x^{(s)}_1,\ldots ,x^{(s)}_d)$. In Figure
\ref{fig:2dim} we have sketched $\left|\omega_{x^{(s)}
y^{(s)}}/K\right|_C$ for a square lattice in two dimensions with
the pacemaker at $\vec{x}^{(s)}=(x^{(s)} ,y^{(s)})$, $x^{(s)}\in
[0,1,...,6]$, $ y^{(s)}\in [0,1,...,6]$, $\alpha=1.0$ (Figure
\ref{fig:2dim}a) and  $\alpha=100.0$ (Figure \ref{fig:2dim}b),
$\epsilon_s=0$.
The largest range for synchronized states is obtained if the
pacemaker is either put in the center of the square (for open
boundary conditions), or on a torus in two dimensions (as
generalization of the closed ring in $d=1$ dimension). The
different shape of the "hats" at the boundaries is due to the
dependence of the critical ratio on $\alpha$, in particular at the
boundaries, as it was visible also in Fig.\ref{fig:alpha_chain}.
Recall that Eq. (\ref{eq:critic_bound}) was derived only for
$\alpha\to\infty$.

\section{Above the Critical Threshold}
\label{sec:above_critical} 
\begin{figure}[ht]
\includegraphics*[width=0.97\textwidth]{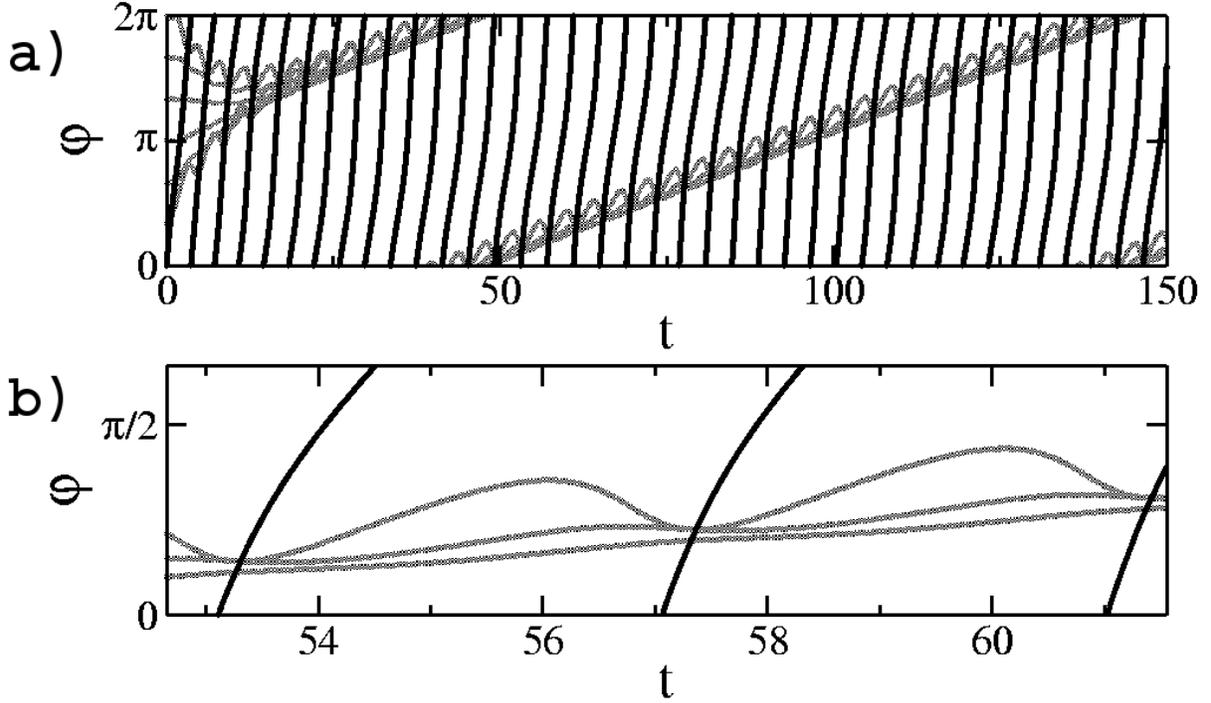}
\caption{Phase portrait above the critical threshold
($\omega/K=2>\left|\omega/K\right|_C$) for  $N=6$ oscillators on a
ring with $\alpha=5.0$ and $\epsilon=0$. The almost vertical black
lines represent the pacemaker's phase, while the phases of the
other oscillators are plotted with thin dark-grey lines. a) The
pacemaker moves faster (steep slope) than the rest of the system
that seems to move coherently with a common frequency; b) zooming
into the phase portrait, it is clear that the phase of each of the
oscillators different from the pacemaker oscillates around the
common frequency with an amplitude inversely proportional to the
distance from the pacemaker.} \label{fig:above}
\end{figure}

\begin{figure}[ht]
\includegraphics*[width=0.97\textwidth]{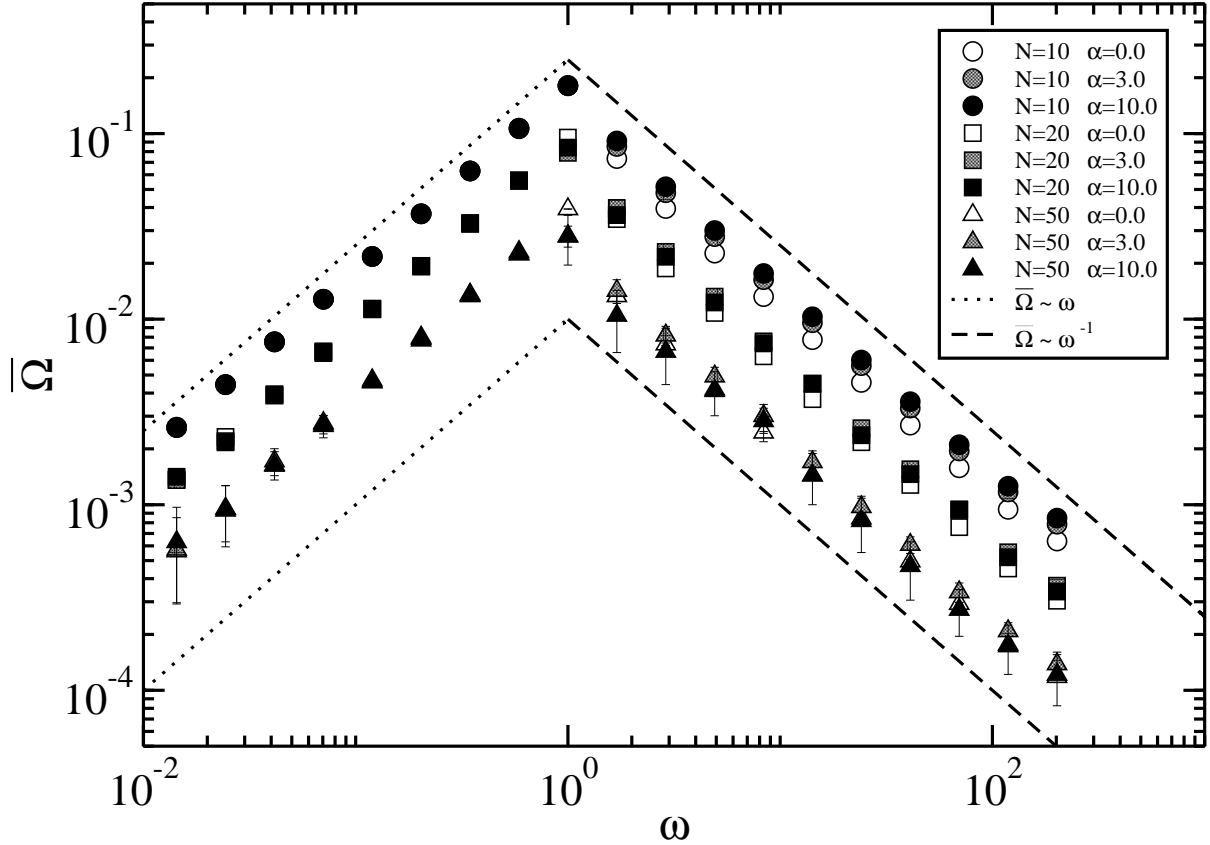}
\caption{Dependence of the average frequency $\overline{\Omega}$
[Eq.(\ref{eq:omega_average})] as  function of the pacemaker's
frequency $\omega$. Numerical results refer to the case of a ring,
with $\epsilon=0$, $\omega>0$, $K=1$  and to different sizes of
the system (i.e. different values of $N$) and/or ranges of the
interactions (i.e. different values of $\alpha$). In this case,
system (\ref{def:model}) is integrated  using the eight-order
Runge-Kutta method with variable time step. Each point is given by
the average over twenty different trials (for which the initial
phases are chosen at random from a uniform distribution) and the
error bars represent the standard deviation from the mean values.
Error bars appear only from $N=50$ on and for $\alpha>0$, because
the value of $t_0 \sim 10^3$ is not always large enough to ensure
that the system reaches the stable state for the given initial
condition.} \label{fig:average_frequency}
\end{figure}

In this section we study the phase
evolution of the oscillators when the ratio of the pacemaker's
eigenfrequency $\omega$ over the coupling $K$ lies above the
threshold of phase-locked motion. As our numerical integration of
the differential equations shows, there are still some regular
remnants to phase-locked motion above the transition. Far above
the transition the oscillators can no longer follow the pacemaker
and get stuck to their eigenfrequency zero. These features are
manifest in a phase portrait $\varphi_i(t)$ of all oscillators $i$
as a function of time, and in some average frequency
$\overline{\Omega}(\omega)$ as a function of the pacemaker's
eigenfrequency $\omega$; $\overline{\Omega}$ replaces the common
frequency $\Omega$ below the critical threshold. Its precise
definition is given below. Here we consider only the
one-dimensional ring-topology. We report our results in Figures
\ref{fig:above} and \ref{fig:average_frequency}.

\subsubsection*{\bf{Phase portrait}} Figure \ref{fig:above}a shows the
portrait of a non-synchronized ring with $N=6$, $\alpha=5.0$,
$\epsilon=0$, and
$\left|\omega/K\right|=2.0>\left|\omega/K\right|_C$. The phases
$\varphi$ are projected to the interval $[0,2\pi)$. The fat
vertical lines represent the phase evolution of the pacemaker,
while the thin dark-grey lines correspond to the phase evolution
of all other oscillators which seem to almost coincide along these
lines, but a zoom, shown in Figure \ref{fig:above}b, displays
their difference in amplitudes as well as the superposition of a
fast and a slow motion. Rewriting the differential equation for
the $i$-th oscillator from system (\ref{def:model}), with periodic
boundary conditions, as
\begin{equation}
\dot{\varphi}_i = \frac{K}{\eta} \frac{\sin{\left(\varphi_s - \varphi_i\right)}}{r_{si}^\alpha} + \frac{K}{\eta} \sum_{j \neq s,i} \frac{\sin{\left(\varphi_j-\varphi_i\right)}}{r_{ji}^\alpha}  \;\;\; , \; \forall \; i \neq s
 \;\;\; ,
\label{eq:fast_motion}
\end{equation}
we can distinguish the interaction between the $i$-th oscillator
and the pacemaker from the interactions between the $i$-th
oscillator and the rest of the system. Calculating from
Eq.(\ref{eq:fast_motion}) the average frequency
$\overline{\Omega}$ of all the oscillators different from the
pacemaker
\begin{equation}
\overline{\Omega}: = \frac{1}{N} \sum_{i \neq s} \dot{\varphi}_i =
\frac{K}{N \eta}\sum_{i \neq s} \frac{\sin{\left( \varphi_s -
\varphi_i \right)}}{r_{si}^\alpha} \;\;\; ,
\label{eq:omega_average_real}
\end{equation}
we see that $\overline{\Omega}$ is given by a weighted average of
the interactions between the pacemaker and the other oscillators,
without any contribution from the mutual interactions between
oscillators different from the pacemaker.
\\
Numerically we approximated the average frequency of Eq.(\ref{eq:omega_average_real}) as
\begin{equation}
\overline{\Omega} = \frac{1}{N}\sum_{i \not = s}
\frac{\varphi_i(t_0+t_1)-\varphi_i(t_0)}{t_1-t_0} \;\;\; ,
\label{eq:omega_average}
\end{equation}
with $10^5 \sim t_1 \gg t_0 \sim 10^3$. The values of $t_1$ and
$t_0$ are chosen sufficiently large to capture only the slow
motion ($2\pi/(t_1-t_0) \ll \omega$). Note that
$\overline{\Omega}$ reduces to $\Omega$ when
$\left|\omega/K\right| \leq \left|\omega/K\right|_C$, since in the
phase-locked regime
$\frac{\varphi_i(t_0+t_1)-\varphi_i(t_0)}{t_1-t_0} = \Omega \;\;\;
, \; \forall \; i$.

\subsubsection*{\bf{Average frequency as a function of the pacemaker's
eigenfrequency}} Moreover, it is of interest to study the absolute
value of  the average frequency $\overline{\Omega}$ as a function
of the pacemaker's eigenfrequency  $\omega$, see Figure
\ref{fig:average_frequency}, where we have chosen $K=1$ for
simplicity as well as $\omega>0$. For $\omega\leq \omega_C \simeq
1$ we obtain the linear increase of $\overline{\Omega} = \Omega
\propto \omega$, as we expect from
Eq.(\ref{eq:common_frequency_ring}). For $\omega\to\infty$ we
expect $\overline{\Omega}\to 0$, since the oscillators can no
longer follow the pacemaker when its eigenfrequency goes to
infinity. However, we have no analytical understanding of the
decay of $\overline{\Omega}$ as an inverse power of $\omega$,
$\overline{\Omega} \propto \omega^{-1}$ for $\omega \gg 1$. The
dotted envelopes mark $\overline{\Omega}\propto \omega$, while the
dashed curves correspond to $\overline{\Omega}\propto
\omega^{-1}$. The numerical results show a weak dependence on the
interaction range ($\alpha$ varies between $0$ and $10$), as well
as a stronger dependence on the size $N$ of the system [for
$\omega \leq \omega_C$ as predicted by
Eq.(\ref{eq:common_frequency_ring})], $N$ varying between $10$ and
$50$. For the numerical integration of the differential equations
we used the eight-order Runge-Kutta algorithm with adapted time
steps in order to accelerate the integration with variable time
step. This introduces some ``noise'' in the numerical estimate of
the first derivatives. Each data point corresponds to an
unweighted average over twenty different choices of the initial
phases (i.e. chosen at random from a uniform distribution). (To
achieve a complete independence on the choice of initial
conditions and a full stabilization would need a longer simulation
time.) The size of the errors increases with $N$ as indicated in
Figure \ref{fig:average_frequency}.

\section{Summary and Conclusions}
\label{sec:conclusions} We considered ensembles of Kuramoto
oscillators with zero eigenfrequency apart from one, the so-called
pacemaker, with eigenfrequency larger than zero and coupled to the
others in a symmetric ($\epsilon_s=0$) or asymmetric ($-1\leq
\epsilon<0$) way. The oscillators were assigned to d-dimensional
regular topologies, in particular to a one-dimensional open chain,
a one-dimensional ring, a Cayley-tree with $z$ branches and a
hypercubic lattice in $d$ dimensions. In general, we studied the
conditions for having phase-locked motion as a function of the
system size $N$, the coupling strength $K$, the pacemaker's
eigenfrequency $\omega_s$ and the asymmetry parameter
$\epsilon_s$. The interaction range between the oscillators was
varied between next-neighbor and all-to-all couplings via the
parameter $\alpha$. We derived the common frequency $\Omega$ of
the phase-locked motion, and the critical threshold
$\left|\omega_s/K\right|_C$, analytically for the limiting cases
$\alpha\to\infty$ and $\alpha=0$, in particular for a chain with
arbitrary position of the pacemaker and a ring in one dimension,
as well as for a Cayley-tree with coordination number $z\geq 2$.
Intermediate coupling ranges $0<\alpha<\infty$ and the phase
evolution above the critical threshold for phase-locked motion
were treated numerically.
\\
In the large $N$ limit, phase-locked motion is impossible in all
cases we have considered, that is, for both chain- and ring-
topologies, and for symmetric $(\epsilon_s=0)$ and asymmetric
$(-1\leq\epsilon_s < 0)$ couplings of the pacemaker. For
$(\epsilon=-1)$, the common frequency $\Omega$ converges to the
pacemaker's $\omega_s$, but $\left|\omega_s/K\right|_C$ goes to
zero for $N\to\infty$, whereas for $\epsilon=0$ $\Omega$ goes to
zero while $\left|\omega_s/K\right|_C$ goes to one for
$N\to\infty$. Similarly, on a Cayley-tree, which we studied for
$\epsilon_0=0$, $\Omega$ and $\left|\omega_0/K\right|_C$ go to
zero for $N\to\infty$, independently of whether the limit is
realized as $z\to\infty$, or $R\to\infty$, or both. However, we
were not only interested in the "thermodynamic" limit and do not
mean the transition to the desynchronized phase for finite $N$ in
the thermodynamic sense of a phase transition. Our main result for
finite $N$ and $-1\leq\epsilon\leq 0$ was that for a whole range
of ratios $\left|\omega_s/K\right|$ it is possible to induce
synchronization by a mere closure of a chain to a ring. (For
ratios below this range the system synchronizes even for an open
chain with pacemaker at position $s$, for ratios above this range
synchronization is neither possible on a ring.)

Two interesting results were obtained from the numerical
integration of the differential equations. Firstly, as a function
of the interaction range (parameterized by $\alpha$), the critical
ratio $\left|\omega/K\right|_C$ showed a minimum at some
intermediate value around $\alpha_M\leq 3$ (for given $N$ and
$\epsilon$ that we considered). This non-monotonic shape was
obtained for our choice of normalization $\eta_i$, defined in
Eq.(\ref{def:eta}). For the pacemaker it is therefore as easy  to
entrain the phases of the other oscillators if it only interacts
with nearest neighbors as if it interacts with all other
oscillators as nearest neighbors (mean-field case). This is well
understandable from Eq.s(\ref{app:next_neighbor1}) and
(\ref{app:next_neighbor2}) in the Appendix, from which it is seen
that it is sufficient for the pacemaker to entrain its nearest
neighbors in order to entrain all other oscillators. Formulas
(\ref{app:next_neighbor1}) and (\ref{app:next_neighbor2}) were
derived under the condition of nearest-neighbor interactions.
Therefore the argument does not go through for intermediate values
of $\alpha$, for which the numerical results show that the
pacemaker has a "harder job" to entrain the remaining ensemble.

Secondly, above the transition from the phase-locked motion to the
desynchronized phase, we found some regular remnants of
phase-locked motion. It is still the pacemaker that determines the
fast frequency of phase fluctuations of the others around some
average value $\overline{\Omega}$ (also determined by the
pacemaker), and it is the distance to the pacemaker that
determines the amplitude of these fluctuations. The value of the
average frequency $\overline{\Omega}$ varies slowly as compared to
the pacemaker's frequency $\omega$. As expected,
$\overline{\Omega}$ approaches zero for
$\left|\omega/K\right|_C\to\infty$ as an expression of the fact
that the system can no longer follow the pacemaker in this limit.
The simple power-law of the decrease of $\overline{\Omega}$
remains to be understood analytically.
\\
The sensitive dependence of synchronization on the topology in a
certain range of parameters may be exploited in artificial
networks and is -very likely- already utilized in natural systems,
in which a switch to a synchronized state should be easily
feasible.

\begin{appendix}

\section*{Appendix}\label{appendix}

\subsection*{Kuramoto Model on a $d$-dimensional lattice}
\label{app_lattice}

We start with the system given by
Eq.s(\ref{def:model}) and (\ref{def:eta}), defined on a
$d$-dimensional lattice. Imposing the
phase-locked condition
\[
\dot{\varphi}_i \equiv \Omega \;\;\; , \; \forall \; i=0,\ldots,N
\; \; \; \; \; \; ,
\]
we have
\[
\Omega = \delta_{i,s} \omega_s + (1+\delta_{i,s}\epsilon_s) \
\frac{K}{\eta_i} \ \Omega_i \; \; \; , \; \forall \ i=0,\ldots, N
\; \; \; ,
\]
where
\[
\Omega_i := \sum_{j \not = i }
\frac{\sin{(\varphi_j-\varphi_i)}}{r_{ji}^\alpha} \;\;\; .
\]

\subsection*{Common frequency in the phase-locked regime}
\label{app_common_frequency} As it is easily seen from the odd
parity of the sine function,
\[
\sum_i \Omega_i = 0 \ \ \textrm{ and } \ \ \sum_{i \neq g}
\Omega_i = -\Omega_g \; ,
\]
so that
\[
\begin{array}{l}
\sum_{i\neq s} \Omega \frac{\eta_i}{K}= \sum_{i\neq s}
\delta_{i,s} \frac{\eta_i\omega_s}{K}+\sum_{i\neq s}
(1+\delta_{i,s}\epsilon_s) \Omega_i
\\
\Rightarrow \ \frac{\Omega}{K}\sum_{i\neq s} \eta_i = - \Omega_s
\end{array}
\; \; \; ,
\]
while for $j\neq s$
\[
\begin{array}{l}
\sum_{i\neq j} \Omega \frac{\eta_i}{K}= \sum_{i\neq j}
\delta_{i,s} \frac{\eta_i\omega_s}{K}+\sum_{i\neq j}
(1+\delta_{i,s}\epsilon_s) \Omega_i
\\
\Rightarrow \ \frac{\Omega}{K}\sum_{i\neq j} \eta_i =
\frac{\eta_s\omega_s}{K}- \Omega_j + \epsilon_s \Omega_s
\\
\Rightarrow \ \frac{\Omega}{K}\sum_{j\neq s}\sum_{i\neq j} \eta_i
= \sum_{j\neq s}\frac{\eta_s\omega_s}{K}- \sum_{j\neq s} \Omega_j
+ \sum_{j\neq s} \epsilon_s \Omega_s
\\
\Rightarrow \ \frac{\Omega}{K}\sum_{j\neq s}\sum_{i\neq j} \eta_i
= \frac{N\eta_s\omega_s}{K}+(1+\epsilon_s N)\Omega_s
\end{array}
\; \; \; ,
\]
which implies
\[
\frac{\Omega}{K}\Big[(1+\epsilon_s N)\sum_{i\neq s} \eta_i +
\sum_{j\neq s}\sum_{i\neq j} \eta_i\Big] =
\frac{N\eta_s\omega_s}{K}\; \; \; .
\]
The  common frequency  $\Omega$ then follows as in Eq.(\ref{eq:common_frequency}).
\\
When the topology has a symmetry such that the normalization
factor defined in Eq.(\ref{def:eta}) is independent of the index
$i$
\[
\eta_i \equiv \eta \;\;\;  , \; \forall \; i=0,\ldots , N \;\;\; ,
\]
then
\[
\left( 1 + \epsilon N \right) \sum_{i \neq s} \eta + \sum_{j \neq s} \sum_{i \neq j} \eta= \left( 1 + \epsilon N \right) N \eta + N^2 \eta = N\eta \left(N+\epsilon N +1\right)\;\;\; ,
\]
thus we obtain Eq.(\ref{eq:common_frequency_ring}).
\\
Such symmetries are realized for $\alpha = 0$, and for arbitrary
$\alpha$ if we have periodic boundary conditions.

\subsection*{Critical ratio $\left|\omega_s/K\right|_C$ for
the one-dimensional lattice} In the case of  next-neighbor
interactions (i.e. for $\alpha \rightarrow \infty$), on an open
chain, we can solve the system (\ref{def:model}) recursively. We
start from the equation for the pacemaker (assuming $0 < s < N$
and $\epsilon_s \neq -1$) and move first to the right of the
pacemaker
\[
\begin{array}{l}
\Omega = \frac{K}{2}\Big[ \sin{(\varphi_{s}-\varphi_{s+1})} +
\sin{(\varphi_{s+2} -\varphi_{s+1})}\Big]
\\
\Leftrightarrow \ \sin{(\varphi_{s+2} -\varphi_{s+1})} =
\frac{2\Omega}{K} + \sin{(\varphi_{s+1}-\varphi_{s})}\;.
\end{array}
\]
For all $\ 1\leq j\leq N-s$ it then follows that
\[
\begin{array}{l}
\Omega = \frac{K}{2}\Big[
\sin{(\varphi_{s+(j-2)}-\varphi_{s+(j-1)})} + \\
\quad + \sin{(\varphi_{s+j}-\varphi_{s+(j-1)})}\Big]
\\
\Leftrightarrow \ \sin{(\varphi_{s+j} -\varphi_{s+(j-1)})} =
\\
=\frac{2\Omega}{K} + \sin{(\varphi_{s+(j-1)}-\varphi_{s+(j-2)})} =
\\
= (j-1)\frac{2\Omega}{K} + \sin{(\varphi_{s+1}-\varphi_{s})}
\end{array}
\; \; \; \; .
\]
In particular for $j=N-s$, we have
\[
\sin{(\varphi_{N} -\varphi_{N-1})} = (N-s-1)\frac{2\Omega}{K}
\sin{(\varphi_{s+1}-\varphi_{s})} \; .\]

Imposing the boundary conditions
\[
\begin{array}{l}
\Omega = K \sin{(\varphi_{N-1} -\varphi_{N})}
\\
\Leftrightarrow \ \sin{(\varphi_{N-1} -\varphi_{N})} =
\frac{\Omega}{K}\;\;,
\end{array}
\]
we obtain for all $ 1\leq j \leq N-s$
\begin{equation}
\sin{(\varphi_{s+j} -\varphi_{s+(j-1)})} =
(2s-2N-1)\frac{\Omega}{K}+ 2j \frac{\Omega}{K}\;\;\; ,
\label{app:critic_right}
\end{equation}
which implies Eq.(\ref{eq:critic_right}) by imposing the condition
that the sine is a bounded function ($|\sin x| \leq 1$). When
$\omega_s>0$ [$\omega_s<0$],  Eq.(\ref{app:critic_right}) is
always negative [positive] and has its minimum [maximum] value for
$j=1$. So if
\begin{equation}
\begin{array}{ll}
\sin{(\varphi_{s+1} -\varphi_{s})} \geq -1 & \textrm{ , if } \omega_s>0
\\
\sin{(\varphi_{s+1} -\varphi_{s})} \leq 1 & \textrm{ , if } \omega_s<0
\end{array}
\Longleftrightarrow \left| \sin{(\varphi_{s+1} -\varphi_{s})} \right| \leq 1\; \; \; , \; \forall \; \omega_s \neq 0
\label{app:next_neighbor1}
\end{equation}
then
\begin{equation}
\begin{array}{ll}
\sin{(\varphi_{s+j} -\varphi_{s+(j-1)})} \geq -1 & \textrm{ , if } \omega_s>0
\\
\sin{(\varphi_{s+j} -\varphi_{s+(j-1)})} \leq 1 & \textrm{ , if } \omega_s<0
\end{array}
\Longleftrightarrow \left| \sin{(\varphi_{s+j} -\varphi_{s+(j-1)})} \right| \leq 1\; \; \; , \; \forall \; \omega_s \neq 0 \; \; \; ,
\label{app:next_neighbor2}
\end{equation}
$\forall \; 1 \leq j \leq N-s$. In other words, if the first
neighbor on the right hand side of the pacemaker (the $(s+1)$-th
oscillator) follows the pacemaker [i.e. if
Eq.(\ref{app:next_neighbor1}) is satisfied], then all the others
follow the pacemaker with the same velocity [i.e.
Eq.(\ref{app:next_neighbor2}) is also satisfied].
\\
In the same way we obtain Eq.(\ref{eq:critic_left}) for
oscillators to the left of the pacemaker. Of course we need that
both the left and the right parts are synchronized in frequency,
hence Eq.(\ref{eq:critic}) is the global solution.
\\
Eq.(\ref{eq:critic_bound}), for which the pacemaker is at the
first ($s=0$) [or at the last ($s=N$)] position of the chain, is
obtained in the same way, but in these particular cases only a
motion to the right [or to the left] is possible because the left
[or the right] part does not exist. In case of the ring the
procedure for deriving Eq.(\ref{eq:critic_ring}) is the same,
apart from applying the different boundary conditions, periodic
ones in this case.

\vspace{0.5cm}

For $\alpha=0$ and phase-locked conditions, the system
(\ref{def:model}) and (\ref{def:eta}) becomes
\[
\Omega = \delta_{i,s}\omega_s +
(1+\delta_{i,s}\epsilon_s)\frac{K}{N}\sum_j
\sin{(\varphi_j-\varphi_i)} \; \; \; .
\]
Introducing the definition of Eq.(\ref{def:order_parameter})
(which serves as an order parameter in a mean-field approximation
that here becomes exact), we write
\[
\left\{
\begin{array}{l}
\Omega=\omega_s+(1+\epsilon_s)KR\sin{(\psi-\varphi_s)}\;\;\; for
\;\;i=s
\\
\Omega=\frac{K}{N}\sin{(\varphi_s-\varphi_i)}+ KR
\sin{(\psi-\varphi_i)} \;\;\; , \; \forall \; i \neq s
\end{array}
\right. \; \; \; \; .
\]
Here all oscillators different from the pacemaker satisfy exactly
the same equation. One possible solution is that all the phases of
the oscillators different from the pacemaker are equal
\[
\varphi_i \equiv \psi \;\;\; ,  \; \forall \; i \neq s \;\;\; ,
\]
from which it follows that
\[
\Omega = \frac{K}{N}\sin{(\varphi_s-\varphi_i)} \;\;\; , \;
\forall \; i \neq s\;\;.
\]

Using the last equation, the fact that the sine-function is
bounded and finally the value of the common frequency, previously
calculated in Eq.(\ref{eq:common_frequency_ring}), it is
straightforward to derive Eq.(\ref{eq:critic_ring}).

\subsection*{Kuramoto Model on a Cayley Tree}
\label{app_cayley}

Consider a node at distance $R$ from the pacemaker, reached along
the path $(s_1,s_2,s_3,\ldots,s_{R-1},s_R)$ from the central node.
Here $s_1=1,\ldots,z$ , while $s_i=1,\ldots,z-1$ , $\forall i
=2,\dots,R$. For this node we can write Eq. (\ref{def:cayley}) as
\[
\begin{array}{l}
\dot{\varphi}_{R,s_1,s_2,\ldots,s_{R-1},s_R} =\Omega = \\
=K \sin{(\varphi_{R-1,s_1,s_2,\ldots,s_{R-1}} -
\varphi_{R,s_1,s_2,\ldots,s_{R-1},s_R})}
\\
\Leftrightarrow \ \sin{(\varphi_{R-1,s_1,s_2,\ldots,s_{R-1}} -
\varphi_{R,s_1,s_2,\ldots,s_{R-1},s_R})} = \frac{\Omega}{K}
\end{array}
\; \; \; \; ,
\]
from which we can see that all  oscillators at distance $R$,
independently on the path along which they are reached from the
center, satisfy the same equation.
\\
Proceeding in an analogous way to before, we can write
\[
\begin{array}{l}
\dot{\varphi}_{R-1,s_1,s_2,\ldots,s_{R-2},s_{R-1}} = \Omega =
\\
= \frac{K}{z} \Big[ \sum_{s_R=1}^{z-1}  \
\sin{(\varphi_{R,s_1,s_2,\ldots,s_{R-1},s_R} -
\varphi_{R-1,s_1,s_2,\ldots,s_{R-1}})} +
\\
\quad + \sin{(\varphi_{R-2,s_1,s_2,\ldots,s_{R-2}} -
\varphi_{R-1,s_1,s_2,\ldots,s_{R-1}})} \Big] =
\\
= \frac{K}{z} \Big[ - (z-1) \frac{\Omega}{K} +
\sin{(\varphi_{R-2,s_1,s_2,\ldots,s_{R-2}} -
\varphi_{R-1,s_1,s_2,\ldots,s_{R-1}})} \Big]
\\
\Leftrightarrow \ \sin{(\varphi_{R-2,s_1,s_2,\ldots,s_{R-2}} -
\varphi_{R-1,s_1,s_2,\ldots,s_{R-1}})} = \frac{\Omega}{K} [ z +
(z-1)]\; \;,
\end{array}
\]
next
\[
\begin{array}{l}
\dot{\varphi}_{R-2,s_1,s_2,\ldots,s_{R-2}} = \Omega =
\\\frac{K}{z} \Big[ \sum_{s_{R-1}=1}^{z-1}  \
\sin{(\varphi_{R-1,s_1,s_2,\ldots,s_{R-2},s_{R-1}} -
\varphi_{R-2,s_1,s_2,\ldots,s_{R-2}})} +
\\
+ \sin{(\varphi_{R-3,s_1,s_2,\ldots,s_{R-3}} -
\varphi_{R-2,s_1,s_2,\ldots,s_{R-2}})} \Big] =
\\
= \frac{K}{z} \Big[ - (z-1) \frac{\Omega}{K} [z+(z-1)] +
\sin{(\varphi_{R-3,s_1,s_2,\ldots,s_{R-3}} -
\varphi_{R-2,s_1,s_2,\ldots,s_{R-3},s_{R-2}})} \Big]
\\
\Longrightarrow \ \sin{(\varphi_{R-3,s_1,s_2,\ldots,s_{R-3}} -
\varphi_{R-2,s_1,s_2,\ldots,s_{R-3},s_{R-2}})} = \frac{\Omega}{K}
[ z + z(z-1)+(z-1)^2]
\end{array}
\]
and so on, until we arrive at
\[
\sin{(\varphi_{R-(t+1),s_1,s_2,\ldots,s_{R-(t+1)}} -
\varphi_{R-t,s_1,s_2,\ldots,s_{R-(t+1)},s_{R-t}})} =
\frac{\Omega}{K}  \ [ \ z \ \sum_{q=0}^{t-1} \ (z-1)^q  \ +  \
(z-1)^t \ ]\; \; \; \; .
\]
For $t=R-1$, we have
\[
\sin{(\varphi_0 - \varphi_{1,s_1})} =  \frac{\Omega}{K}  \ [ \ z \
\sum_{q=0}^{R-2} \ (z-1)^q  \ +  \ (z-1)^{R-1} \ ]  \; \; \; \; ,
\]
but also, at the center of the Cayley tree,
\begin{equation}
\Omega = \dot{\varphi}_0 = \omega_0 +  \frac{K}{z} \ \sum_{s_1=1}^z
\ \sin{(\varphi_{1,s_1} - \varphi_0)} =
 \omega_0 -  K \ \frac{\Omega}{K}  \ [ \ z \ \sum_{q=0}^{R-2} \ (z-1)^q  \ +  \ (z-1)^{R-1} \ ]
\;\;\; .
\label{app:cayley_tree1}
\end{equation}
Eq.(\ref{eq:common_frequency_cayley}) is easily
obtained from Eq.(\ref{app:cayley_tree1}).
\\
As we have seen,  all  oscillators at the same distance $r$ from
the pacemaker satisfy the same equation. We can write
\[
\sin{(\varphi_{r} - \varphi_{r-1})} \ = \  - \ \frac{\omega_0}{K}
\ \frac{z \ \sum_{q=0}^{R-r-1} \ (z-1)^q \ + \ (z-1)^{R-r}}{\ 1 \
+ \  z \ \sum_{q=0}^{R-2} \ (z-1)^q \ + \ (z-1)^{R-1}}\; \; \; \; ,
\]
suppressing the path that was followed to reach the node. It is
convenient to rewrite this equation according to
\[
\begin{array}{l}
z \ \sum_{q=0}^{R-r-1} \ (z-1)^q \ + \ (z-1)^{R-r} =
\\
=\frac{1}{(z-1)^{r-1}} \Big[ z \sum_{q=0}^{R-r-1}(z-1)^{q+r-1}+
(z-1)^{R-r+r-1} \Big] =
\\
= \ \frac{1}{(z-1)^{r-1}} \Big[z \sum_{s=r-1}^{R-2}  \ (z-1)^{s} \
+ \ (z-1)^{R-1} \Big] \ =
\\
= \ \frac{1}{(z-1)^{r-1}} \Big[ z\sum_{s=0}^{R-2}  \ (z-1)^{s} \ -
\ z\sum_{s=0}^{r-2}  \ (z-1)^{s} \ + \ (z-1)^{R-1} \Big] \ =
\\
= \ \frac{1}{(z-1)^{r-1}} \Big[ 1 \ + \ z \sum_{s=0}^{R-2}  \
(z-1)^{s} \  + \ (z-1)^{R-1} \ - \ z\sum_{s=0}^{r-2}  \ (z-1)^{s} \
- \ 1 \Big]
\end{array}
 \ \ . \]
for $R \geq 2$. Finally we obtain
\begin{equation}
\sin{(\varphi_r - \varphi_{r-1})} \ = \frac{\omega_0}{K \
(z-1)^{r-1}} \ \Big[ \ \frac{1 \ + \ z \ \sum_{s=0}^{r-2} \
(z-1)^s}{ 1 \ + \ z \ \sum_{s=0}^{R-2} \ (z-1)^s \ + \
(z-1)^{R-1}} \ - \ 1 \ \Big] \;\;. \label{app_sol_1}
\end{equation}
When $\omega_0>0$, Eq. (\ref{app_sol_1}) is always negative,
monotone and increasing and takes its minimum value for $r=1$,
while when $\omega_0<0$,  Eq. (\ref{app_sol_1}) is always
positive, monotone and decreasing and takes its maximum value for
$r=1$, namely
\[
\sin{(\varphi_1-\varphi_0)} \ = \ \frac{\omega_0}{K} \ \Big[ \
\frac{1}{ 1 \ + \ z \ \sum_{s=0}^{R-2} \ (z-1)^s \ + \
(z-1)^{R-1}} \ - \ 1 \ \Big] \; \; \; \; .
\]
In order to derive the maximal value for
$\left|\omega_0/K\right|$, we use the  bound on sine function,
from which we find the critical fraction of
Eq.(\ref{eq:critic_cayley}).

\end{appendix}

\end{document}